\documentclass[preprint]{aastex}

\usepackage{graphicx}
\usepackage{natbib,twoopt}
\bibpunct{(}{)}{;}{a}{}{,} 
\usepackage{color}

\newcommand{\dtdtm}{$\left(\Delta T/\Delta t\right)_{\mathrm{max}}$}
\newcommand{\dtdt}{$\left(\Delta T/\Delta t\right)$}
\newcommand{\SIu}{J\,m$^{-2}$s$^{-1/2}$K$^{-1}$}

\slugcomment{Published in ApJL Vol. 810, Issue 2, L22, 5p.}

\shorttitle{Rapid temperature changes and early activity on comet 67P}
\shortauthors{Al\'i-Lagoa et al.}

\begin{document}

\title{Rapid temperature changes and the early activity on comet 67P/Churyumov-Gerasimenko}

\author{
  V. Al\'i-Lagoa$^\dagger$, 
  M. Delbo',
  and 
  G. Libourel
}
\email{$^\dagger$valilagoa@oca.eu}

\affil{
  Laboratoire Lagrange, Universit\'e C\^ote d'Azur, Observatoire de la C\^ote d'Azur, CNRS \\Blvd de l'Observatoire, CS 34229, 06304 Nice cedex 4, France
}

\begin{abstract}
  The so-called ``early activity'' of comet 67P/Churyumov-Gerasimenko has been observed to originate mostly in parts of the concave region or ``neck'' between its two lobes. Since activity is driven by the sublimation of volatiles, this is a puzzling result because this area is less exposed to the Sun and is therefore expected to be cooler on average \citep{Sierks2015}.
  
  We used a thermophysical model that takes into account thermal inertia, global self-heating, and shadowing, to compute surface temperatures of the comet. We found that, for every rotation in the August--December 2014 period, some parts of the neck region undergo the fastest temperature variations of the comet's surface precisely because they are shadowed by their surrounding terrains. 
 Our work suggests that these fast temperature changes are correlated to the early activity of the comet, and we put forward the hypothesis that erosion related to thermal cracking is operating at a high rate on the neck region due to these rapid temperature variations. 
This may explain why the neck contains some ice --as opposed to most other parts of the surface \citep{Capaccioni2015}-- and why it is the main source of the comet's early activity \citep{Sierks2015}.

 In a broader context, these results indicate that thermal cracking can operate faster on atmosphereless bodies with significant concavities than implied by currently available estimates \citep{Delbo2014}. 
\end{abstract}


\section{Introduction \label{sec:intro}}

Comet 67P/Churyumov-Gerasimenko (hereafter 67P), the target of the European Space Agency's Rosetta mission, is a bilobed body \citep{Thomas2015} with a low geometric visible albedo of 0.069 $\pm$ 0.002 and relatively uniform spectral colors. These are typical of a dehydrated, organic-rich carbonaceous surface, although some areas that have been observed to be active show evidence of the presence of water ice \citep{Capaccioni2015}. 

The study of the activity of 67P in the June--September 2014 period, which is referred to as early activity \citep{Gulkis2015,Sierks2015}, has revealed that it originates mostly from parts of the ``neck'' region. In this paper, ``neck'' will refer to the concave part of the surface between the two main lobes that form the body \citep{Thomas2015}. This includes the Hapi and Anuket regions and, possibly, parts of the Hathor and Seth regions. 
This is an unexpected result because the neck is frequently shadowed as the comet rotates around its axis and is thus expected to be colder on average. For example, \citet{Sierks2015} calculated that the neck received 15\% less energy per rotational period than did other parts of the surface on 6 August 2014, and about half as much throughout the whole orbit. 
Sierks et al. point out that the early activity coming from the neck may be related to the fact that the neck region may be structurally or compositionally different, as suggested by its colorimetry differing from that of the rest of the surface \citep{Capaccioni2015}. They also note that a more detailed thermophysical model, one that uses thermal inertia and takes sublimation into account, should also be used to tackle the question. 

In this work, we present and discuss a hypothesis that may explain the characteristics of the early activity of 67P: thermal cracking of surface materials is operating faster on the neck because of the fast changes in temperature resulting from the shadows cast on it by its surrounding terrains. 

Thermal fatigue refers to the damage caused on a solid material when it is forced to undergo periodical temperature variations. This process is understood in terms of the progressive growth of pre-existing cracks within the material. In the case of small solar system bodies, thermal fatigue was invoqued by \citet{Dombard2010} to explain the formation and particular distribution of broken-up rocks near boulders in ``regolith ponds'' on asteroid (433) Eros. \citet{Delbo2014} performed experiments on asteroidal analog materials to quantify the crack growth rate in their samples and showed that some small fragments were separated as the lengths of cracks increased gradually over time. 
This erosion or comminution of the material is referred to as thermal cracking. 
Delbo' et al. used their results to constrain a thermomechanical model to estimate the erosion rate on the surfaces of near-Earth asteroids (NEAs). They concluded that thermal cracking is capable of producing regolith on the surface of NEAs orders of magnitude faster than do micrometeorite impacts, the traditionally considered mechanism.  

The rate of erosion by thermal fatigue is proportional to the temperature excursion ($\Delta T$), i.e., the difference between the maximum and minimum temperatures over each cycle \citep{Delbo2014}. 
In addition, other experiments have shown that the intensity of thermal cracking is increased when rapid temperature variations --high $\left(dT/dt\right)$-- are applied on different materials (e.g., see \citealp{Richter1974}, \citealp{Lu1998}, and references therein). Based on this, we propose that the shadowing of the neck may be increasing its erosion rate, thus exposing subsurface ices faster than in other regions. This could explain both the detection of a small amount of water ice in the neck \citep{Capaccioni2015} and why it is the region from which most of the early activity originated. 

Therefore, our aim is to study the temperatures and their corresponding rates of change over the surface of 67P by means of a thermophysical model (TPM) in order to study this hypothesis' feasibility. 
In addition to thermal inertia, our model takes into account shadowing due to shape concavities as well as self-heating between interfacing parts of the surface (see Section~\ref{sec:TPM} for more details).
Our results, discussion, and conclusions are presented in Sections~\ref{sec:results}, \ref{sec:discussion}, and \ref{sec:conclusions}. 

\section{Thermophysical modelling \label{sec:TPM}}

TPMs are used to calculate the surface and subsurface temperatures of an airless body. They are based on the application of the energy conservation principle to characterise the energy budget (incident, absorbed and reflected radiation) at a given surface element at a given time. The surface temperature distribution of a particular body at a given time does not only depend on its position relative to the Sun and its surface reflectivity or albedo, but also on its shape, its instantaneous spatial orientation --which depends on its rotational state-- and several other relevant physical properties, such as the thermal inertia, or the degree of macroscopic roughness of the surface material \citep[for a review see, e.g.,][]{Harris2002}.

Here we used an upgraded version of the thermophysical model implementation of M. Delbo' \citep{Delbo2007,Ali-Lagoa2014} that incorporates the effects of shadowing and self-heating. Although the model has been mostly applied to non-active asteroids, given the small percentage of water ice on the surface of 67P \citep{Capaccioni2015}, we assume it is appropriate for our purposes of studying the surface temperature distribution when the comet is sufficiently far from the Sun (see Section~\ref{sec:discussion} for a discussion). Next, we give a brief description of the model, and the reader is referred to the literature for more details. 

We take a three-dimensional surface divided into triangular facets as well as the object's rotational properties --the axis' orientation and the period-- as input (see the next section). The implementation of shadowing and self-heating requires the identification and labelling of each facet's landscape, i.e., the set of all other facets visible from its centre, before any calculation of the temperatures. To that end, we first identify all other (\emph{remote}) facets that lie above each (\emph{local}) facet's plane or horizon. Next, we use ray tracing to determine whether the local facet is visible from each one of these remote facets. Once this first step is done, we are ready to use TPM to calculate the temperatures. 

At each time step of the run, we need to determine whether facets are illuminated or not. Even if the Sun is above a facet's horizon, it may still shadowed by one of its landscape facets. Again, we use ray tracing: we take the sun versor and find the parallel line that passes through the local facet's centre; we then calculate, one at a time, the intersection of this line with the plane of each landscape facet; if said intersection lies inside the area of the landscape facet \citep[we use the criterion of][]{Rozitis2011}, we consider the local facet to be shadowed.

Then, the temperature of illuminated facets will depend on the heliocentric distance $r$ and spatial orientation of the object, since the amount of solar energy incident on an illuminated facet is inversely proportional to $r^2$ and proportional to the cosine of the angle subtended between the facet's normal vector and the sunward direction.
  Eventually, a fraction $(1-A)$ of that energy is absorbed, where $A$ is the bolometric Bond albedo. 
  In addition, facets will also absorb infrared radiation emitted by their landscape facets, which is referred to as self-heating. 

Finally, the surface temperatures will also depend on the rate at which the absorbed energy is conducted to subsurface layers \citep[according to the one-dimensional heat diffusion equation and the usual boundary conditions; see][]{Delbo2007}. This is governed by the value of the thermal inertia ($\Gamma$), which is a measure how much a material's temperature response is delayed with respect to changes on its incident energy. Assumed to be constant, $\Gamma$ is defined as the square root of the density, the specific heat capacity, and the conductivity of the material.

\subsection{Model input parameters}
Given the comet's low visible albedo, we did not consider the effect of second- or higher-order reflected sunlight, and we assumed that the infrared albedo is zero \citep[see, e.g.,][]{Rozitis2011}. 
We did not consider the effects of macroscopic surface roughness --usually modelled as self-heating within a hemispherical crater located in each facet-- either, but we discuss its possible effects on our results in Section~\ref{sec:discussion}. 

We used the period and rotational axis orientation given in \citet{Sierks2015} and assumed they are constant. We downloaded the ``SHAP2P'' shape model from ESA's website\footnote{http://sci.esa.int/science-e/www/object/doc.cfm?fobjectid=54726.}, which features more than 60\,000 triangular facets. This is a preliminary shape that does not capture all the details of the surface of the comet and, in particular, the neck region presents some artifacts. Moreover, since it is partially incomplete because the southern part of the comet had remained non-illuminated so far, we only concentrate on the northern side. Given these circumstances, we recomputed a new shape with 16000 facets to significantly reduce our computing times. To ensure that the shape remeshing did not bias our results, we also tested shapes with decreasing resolution (2000, 4000 and 8000 facets) and our conclusions did not change. 

We took a thermal inertia value of $\Gamma = 50$ \SIu, the maximum within the range reported by \citet{Gulkis2015}. This value is characteristic of insulating (low-thermal inertia) fine-grained carbonaceous materials. We chose the maximum as a way of being conservative, in the sense that the temperature changes will be more delayed and the effect of shadows over thermal cracking may be somewhat slowed down. 
In accordance with its low visible geometric albedo, we took a constant bolometric Bond albedo of $A=0.01$. Finally, we assumed a constant value of $\epsilon=0.9$ for the emissivity, which is typical of asteroidal regoliths \citep[see][and references therein]{Delbo2007}. 

\section{Results \label{sec:results}} 

\begin{figure}[!h]
  \begin{center}
  \includegraphics[width=0.48\textwidth]{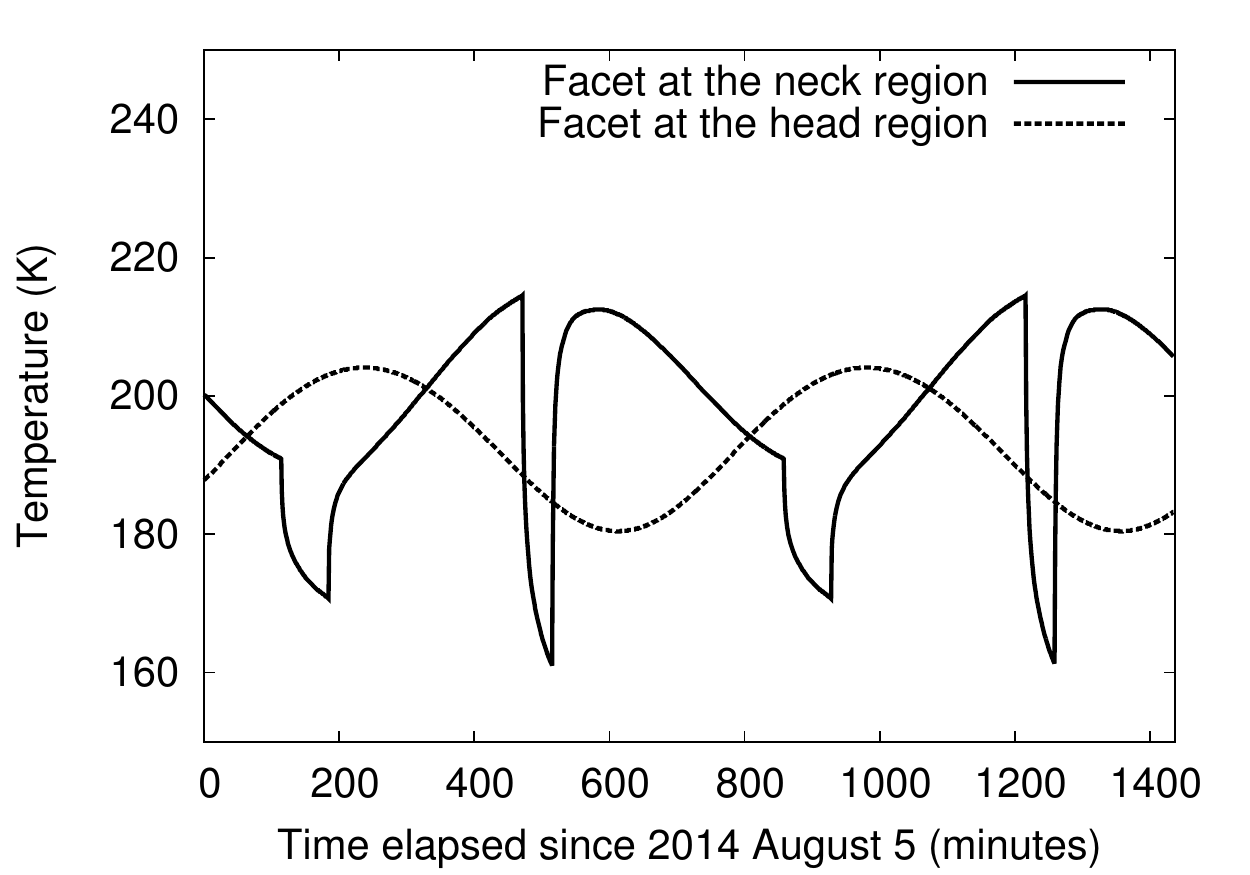}  
  \caption{
    Temperature variation over 24 hours (almost two rotational periods) for two facets. The one located at the head shows a smooth temperature variation (dashed line), whereas the one in the neck region  undergoes two sudden decreases in temperature related to its shadowing by its two neighbouring walls (solid line). 
    \label{fig:DTDt_facets}
  }
  \end{center}
\end{figure}

We calculated the surface temperatures of 67P every minute during 24 hours starting at four different epochs: 00:00:00 UT 5 August, 22 September, 22 November, and 31 December 2014; from now on, these groups will be referred to as epochs 1 through 4. These were chosen based on the work of J. B. Vincent et al. (private communication), who identified the source regions of the observed jets on those dates. 
Figure~\ref{fig:DTDt_facets} shows the temperature evolution as a function the time elapsed since epoch 1 for two facets, one located on the neck region, the other on a convex area of the head. 
The smooth temperature variations of the latter contrast with the two rapid decreases per period experienced by the former, in particular the one ocurring after $\approx$500 minutes. As we expected, these fast temperature changes are the result of the neck region facet being shadowed by its surrounding elevated terrains.  

The two cases shown in Figure~\ref{fig:DTDt_facets} represent the two ends of a continuous range of possibilities. From epochs 1 to 4, the temperature excursions of the north-side facets increase from $\Delta T \sim$ 60 K in epoch 1 to $\sim$90 K in epoch 4. This means that, for most of these facets, the $\Delta T$ are not qualitatively very different in the neck region relative to other parts of the surface.

\begin{figure*}
  \begin{center}
    \includegraphics[width=0.45\textwidth]{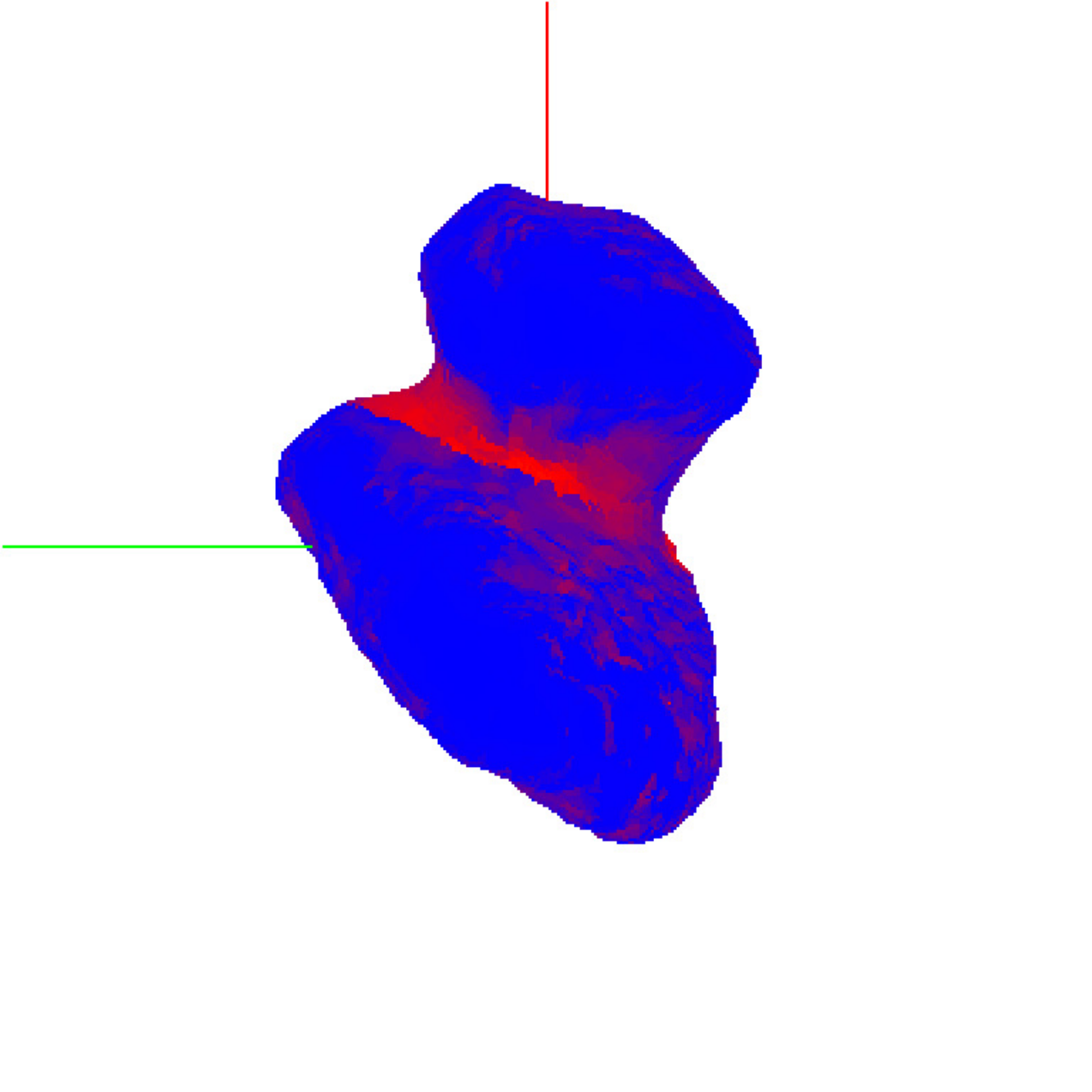}
    \put(0,-8){(a)}
    \includegraphics[width=0.45\textwidth]{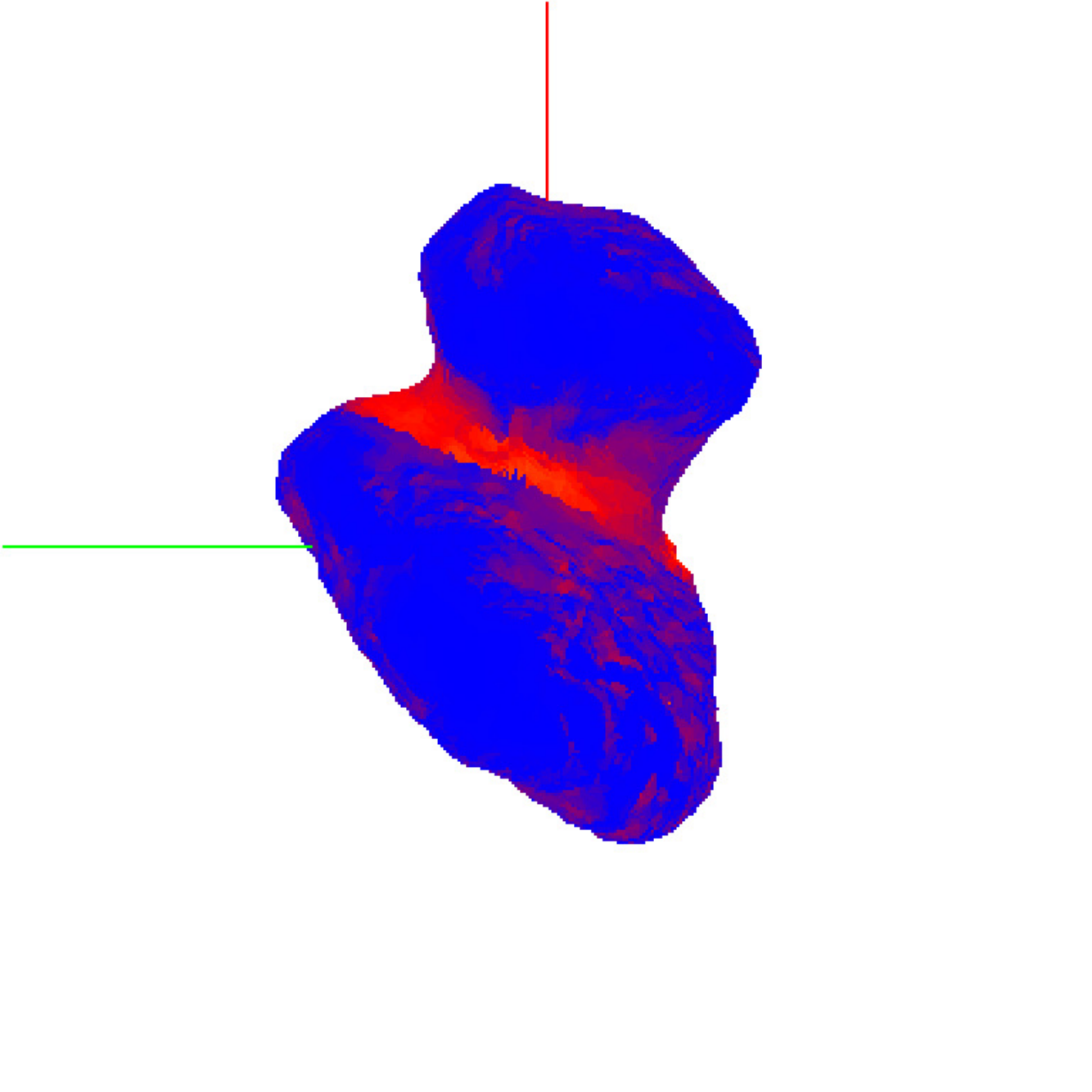}
    \put(0,-8){(b)}
  
    \includegraphics[width=0.45\textwidth]{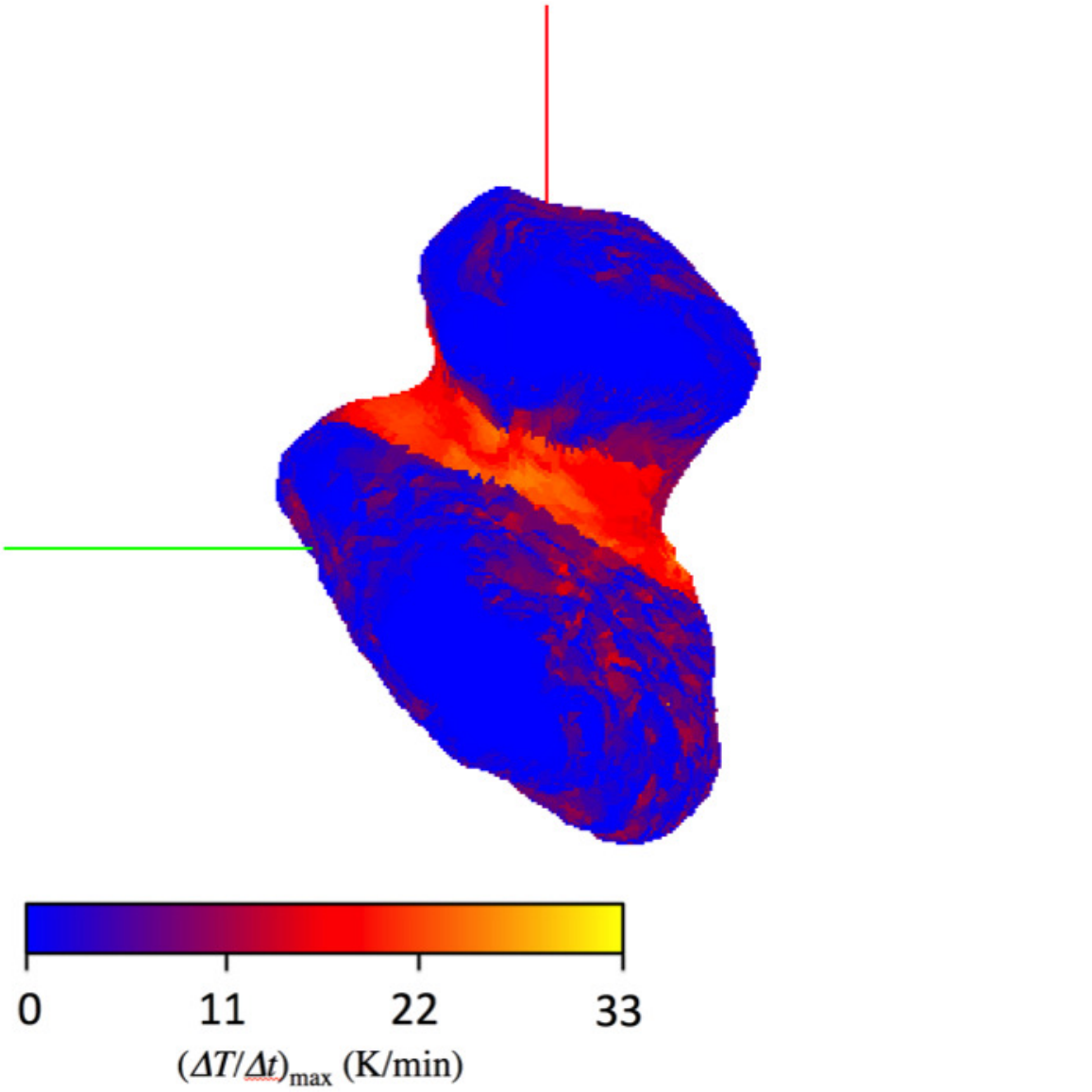}
    \put(0,-8){(c)}
    \includegraphics[width=0.45\textwidth]{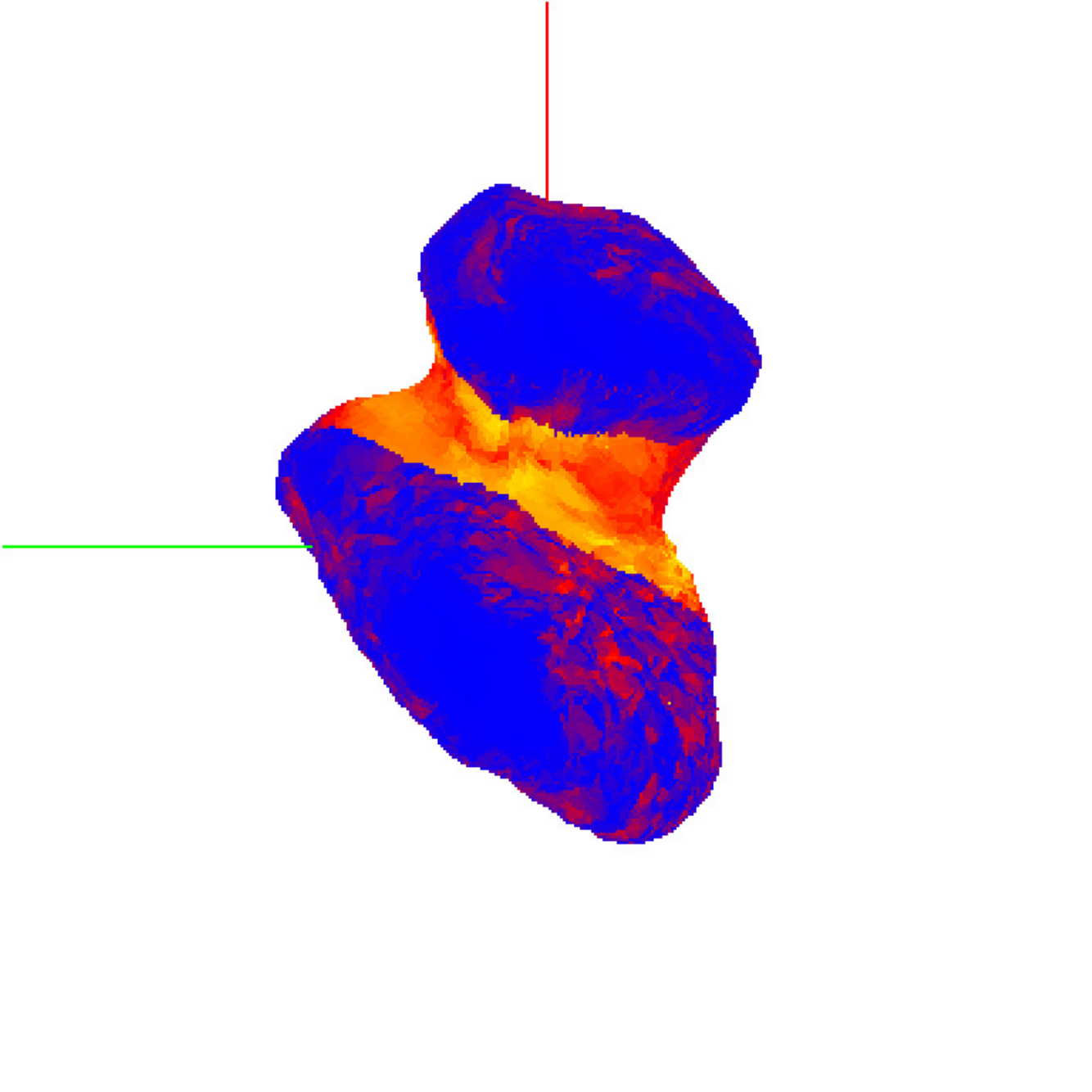}
    \put(0,-8){(d)}
  \caption{
    Maps of \dtdtm\, at the four selected epochs ordered in panels (a) through (d). The comet is viewed from its north pole, and the red and green lines correspond to the $x$ and $y$ axes. The highest values in each plot correspond to 18, 21, 26, and 32 K/min, respectively.
    \label{fig:DTDt}
  }
  \end{center}
\end{figure*}

\begin{figure*}[h!]
  \begin{center}
    \includegraphics[width=0.80\textwidth]{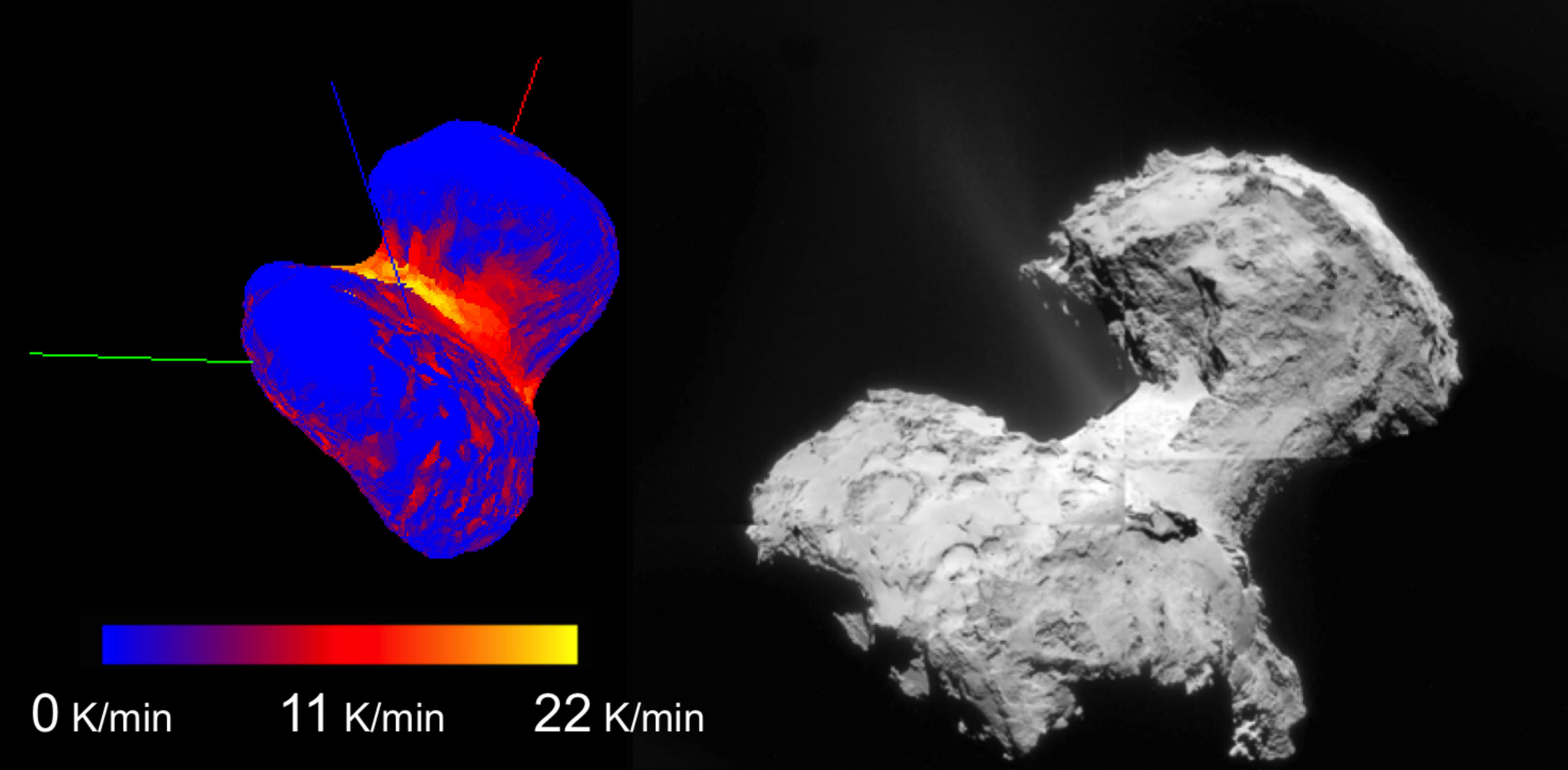}   
    \caption{
      Comparison between our \dtdtm\, map for epoch 2 and an image of 67P taken in 2 September 2014 (image credit ESA/Rosetta/Navcam/Bob King). 
      \label{fig:DTDt_vs_activity}
    }
  \end{center}
\end{figure*}

On the contrary, the temperature rate of change is strikingly different on the neck. Figure~\ref{fig:DTDt} shows the distribution of the maximum instantaneous rate of change of temperature per period, \dtdtm, at epochs 1 through 4. 
The neck region clearly stands out relative to other parts of the surface, which confirms the expected effects of the shadowing on the temperature rate of change (see Section~\ref{sec:intro}). 
The correlation between our high-\dtdtm\, regions and those where activity has been observed to originate is illustrated in Figure~\ref{fig:DTDt_vs_activity}, where we compare an image of the comet taken on 2 September 2015 and our epoch 2 map. For additional comparison we refer the reader to Figure 4 of \citet{Sierks2015} and ESA's website. 

\begin{figure}[h!]
  \begin{center}
    \includegraphics[width=0.50\textwidth]{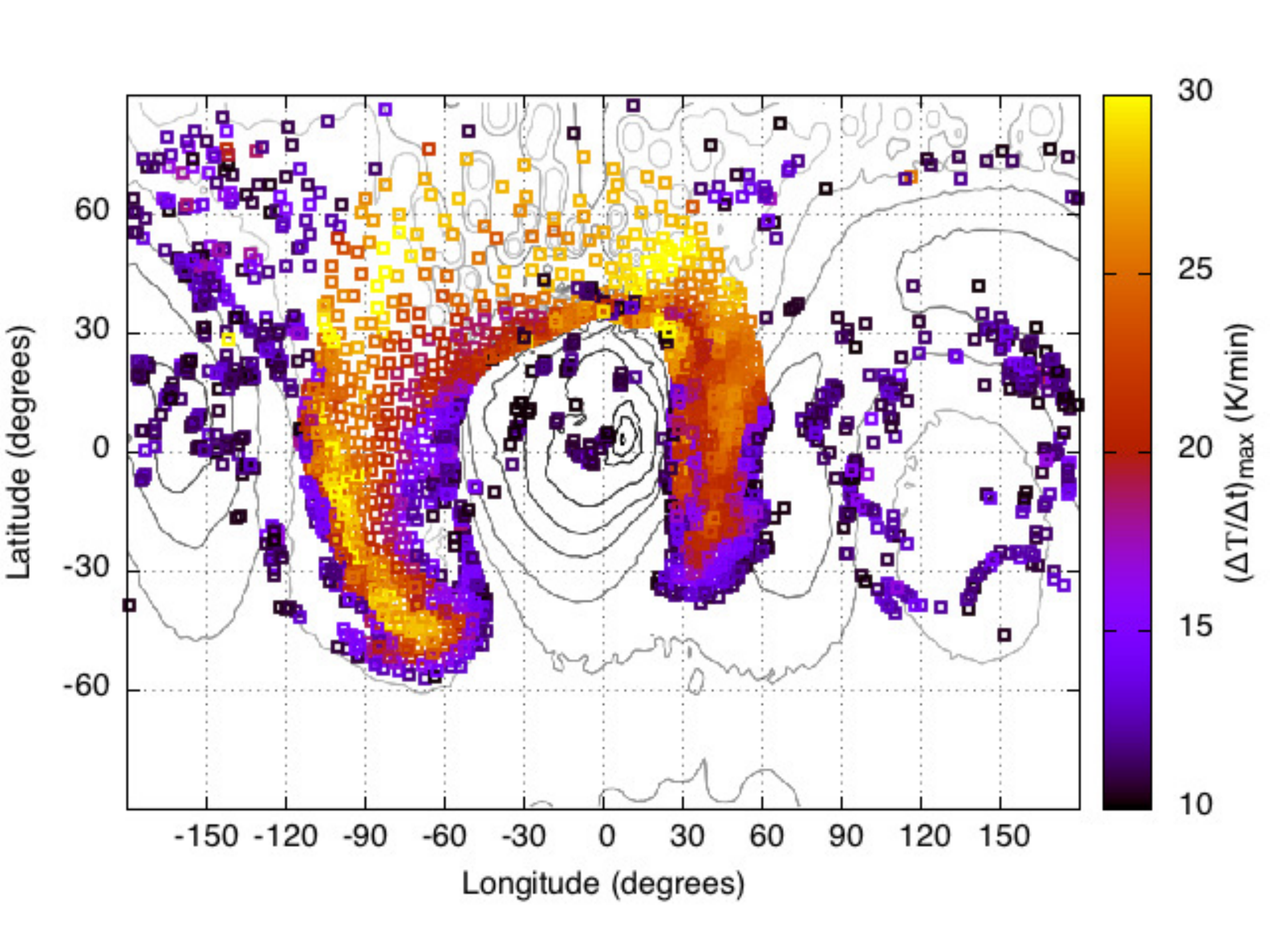}

    \caption{Values of \dtdtm\, computed for epoch 4 and projected onto a topographic map of the shape model's surface. The contour lines fade towards lower heights. Height is calculated as the normalised distance to the origin of coordinates. The areas with highest \dtdtm\, correspond to highly concave parts, especially the neck region. \label{fig:lat_long}}
  \end{center}
\end{figure}

Moreover, we find other smaller regions in the head and the body with moderately high values of \dtdtm\, as well that may correspond to other reported secondary sources of activity \citep{Sierks2015,Vincent2015}. 
These can be better visualised in Figure~\ref{fig:lat_long}, which shows the areas with highest \dtdtm\, per rotational period on epoch 4 projected onto a topographic map. Coloured as a function of each facet's \dtdtm\, at epoch 4, each point's position is given by each facet's barycentre's latitude and longitude. 

\section{Discussion \label{sec:discussion}}

The correlation between the high \dtdtm-regions shown in Figure~\ref{fig:lat_long} and reported active locations \citep{Sierks2015,Vincent2015} constitutes a strong hint that thermal cracking is taking place on concave areas due to the effects of topographic shadows. For the moment, however, we cannot estimate reasonable erosion rates and timescales based solely on our results and the work of \citet{Delbo2014}. The most important obstacle is that relevant laboratory constraints for the thermomechanical model of \citet{Delbo2014} are not yet available. Not only there are no experiments performed on any material that may be representative of a still poorly-characterised cometary surface, but Delbo et al.'s experiments did not explore the effects of rapid temperature variations on the crack growth. 
Progress in these directions would be required. 

In particular, the thermal inertia, the most critical parameter governing the rate of change of the temperature in our model, is not yet well constrained for different types of terrains on 67P. There is a very wide range of possible values of conductivities for water-ice depending on particle sizes and other physical properties \citep{Gundlach2012comet}. If we take a density of 1000 kg\,m$^{-1}$, the heat capacity of carbonaceous chondrites (560 J\,kg$^{-1}$K$^{-1}$), and we broadly vary the conductivity from 0.001 to 0.1 W\,m$^{-1}$K$^{-1}$,  thermal inertia would range from 20 to 200 \SIu. Using these limiting values for epoc 1, for example, leads to maximum values of \dtdtm\, between 40 K/min and 5 K/min, respectively, i.e., about a factor of 2 larger and a factor of 3 smaller than the 18 K/min obtained with $\Gamma=50$ \SIu. Since this value of $\Gamma$ corresponds to a conductivity closer to the low-value end of the possible range ($\approx$0.004 W\,m$^{-1}$K$^{-1}$), our \dtdtm-values should be more typical of low-thermal inertia terrains. In spite of that, our model is reasonably reproducing the typical temperatures measured and modelled by authors of the Rosetta mission's scientific teams \citep[e.g. ][]{Capaccioni2014_abstract,Tosi2015}, and our minimum \dtdtm\, for August 2014 compares well with the minimum estimate of 4 K/min reported by \citet{Capaccioni2014_abstract}. 

This also means that our other simplifying assumptions are not expected to have a significant impact on our results. For example, 
we do not account for multiple-scattered reflected sunlight, which should reduce the maximum \dtdt\, slightly since it increases the amount of energy that shadowed facets receive from illuminated parts of their landscape. But given the comet's low Bond albedo, this effect might be negligible. 
Also, as \citet{Sierks2015} and \citet{Gulkis2015}, we do not include the effects of sublimation in our TPM. This would only reduce the temperatures of the involved facets once they become active and, therefore, it does not argue against the hypothesis put forward here. 

Thus, keeping these caveats in mind, we can still discuss two possible scenarios for the erosion rate of 67P. On the one hand, if it were sufficiently fast, the activity might occur instantly or shortly after the sudden changes in temperature. 
On the other hand, if the erosion rate were slower, thermal craking might still be taking place all throughout the quiescent phases of the comet's orbit, i.e., when it is sufficiently far from perihelion. In this case, ice would progressively be exposed on the concave parts and would subsequently be available for sublimation when the temperatures increase sufficiently as the comet approaches perihelion once again. 
Also note that the maximum stress in a material during a temperature cycle might not necessarily occur at the time of maximum \dtdt\, \citep{Molaro2015}, which might introduce a delay between the fast temperature changes and the activity. 

A fast rate of erosion could also have implications on the comet's morphology, since the neck might be the result of a strong feedback effect operating on an initially shallower concavity: as the erosion proceeds, the slope of the concavity increases so that the values of \dtdtm\, will also increase; this would result in a faster erosion, which would then increase the slope faster, etc. Although this does not necessarily contradict other possible scenarios proposed for the origin of 67P, in which it is considered to be the result of a merger between two bodies \citep{Rickman2015,Jutzi2015}, fast thermal cracking on a pre-existing concavity might also constitute a completely alternative explanation for the comet's morphology.  

Finally, at a broader and more speculative level, faster thermal cracking in concavities might also explain why water ice was detected on asteroids (24) Themis and (67) Cybele \citep{Campins2010,Licandro2011} at heliocentric distances at which water ice is not expected to be stable, and possibly even some of the activated asteroids \citep[see, e.g., ][]{Jewitt2012_AA}. If our hypothesis is not confirmed for the particular case of 67P, regolith production through faster thermal cracking related to shadowing may be operating on many other airless bodies that have significant concavities. 

\section{Conclusions \label{sec:conclusions}}

We used a thermophysical model to study the hypothesis that thermal cracking is eroding the neck region of 67P faster than on other parts of its surface as a consequence of the shadows cast on it by its neighbouring terrains. Our model indeed shows how the shadows cause large, sudden drops in temperature over the neck region (Figure~\ref{fig:DTDt_facets}) which, according to experiments \citep[e.g., ][]{Lu1998}, implies a faster erosion driven by thermal cracking. 
Our results provide the best correlation found so far between the source regions of early activity of 67P and temperature-related effects (Figures~\ref{fig:DTDt} and \ref{fig:lat_long}), and suggest that the shadowing of deep concave parts of atmosphereless bodies may cause thermal-cracking erosion to be much faster than previously estimated \citep{Delbo2014}. This includes both near-Earth objects and main belt asteroids at longer heliocentric distances. 

\acknowledgments

We thank J. B. Vincent, J. Wilkerson, and J. Blum for fruitful discussions. 
We acknowledge support from the France's ANR's project SHOCKS.

\end{document}